\documentclass[twocolumn,aps,a4paper,showpacs,amsfonts,amsmath,amssymb]{revtex4}
\usepackage{graphicx}
\begin{document}

\title{Scaling and data collapse for the mean exit time of asset prices}
\author{Miquel Montero}
\email[Corresponding author: ]{miquel.montero@ub.edu}
\author{Josep Perell\'o}
\author{Jaume Masoliver}
\affiliation{Departament de F\'{\i}sica Fonamental, Universitat de
Barcelona,\\  Diagonal 647, E-08028 Barcelona, Spain}
\author{Fabrizio Lillo,$^{1,2,3}$ Salvatore Miccich\`e,$^{1,2}$ and Rosario N. Mantegna$^{1,2}$}
\affiliation{$~^{1}$ Dipartimento di Fisica e Tecnologie Relative, University of Palermo, Viale delle Scienze, Edificio 18, I-90128 Palermo, Italy\\$~^{2}$ INFM-CNR, Unit\`a di Palermo, Palermo, Italy\\$~^{3}$ Santa Fe Institute, 1399 Hyde Park Road, Santa Fe, NM 87501, USA}

\begin{abstract}
We study theoretical and empirical aspects of the mean exit time of financial time series. The theoretical modeling is done within the framework of continuous time random walk. We empirically verify that the mean exit time follows a quadratic scaling law and it has associated a pre-factor which is specific to the analyzed stock. We perform a series of statistical tests to determine which kind of correlation are responsible for this specificity. The main contribution is associated with the autocorrelation property of stock returns. We introduce and solve analytically both a two-state and a three-state Markov chain models. The analytical results obtained with the two-state Markov chain model allows us to obtain a data collapse of the 20 measured MET profiles in a single master curve.
\end{abstract}
\date{\today}
\maketitle

\section{Introduction}

The Continuous Time Random Walk (CTRW) formalism introduced four decades ago by Montroll and Weiss~\cite{montrollweiss} has been successfully applied to a wide and diverse variety of physical phenomena over the years \cite{weissllibre} but only recently to finance 
\cite{general1,general2,general3,general4,general5,general6,ivanov,masoliver1,masoliver2,montero}. In this latter context, the efforts have been mostly focused on the statistical properties of  the waiting time between successive transactions and the asset return at each transaction. Different studies in different markets are conceiving the idea that the empirical distributions of both random variables are compatible with an asymptotic fat tail behavior \cite{general1,general2,general3,general4,general5,general6,masoliver1,masoliver2,ivanov}. 

Within the CTRW formalism some of us have recently investigated the mean exit time (MET) of asset prices out of a given  interval of size $L$ for financial time series \cite{montero}. This study shows that the MET follows a quadratic growth in terms of $L$ for small interval lengths $L$. In the same study, this functional form was observed for a specific time series of the foreign exchange (FX) market, which is the U.S. dollar/Deutsche mark futures time series \cite{montero}.

In this paper we investigate both theoretically and empirically the MET of price returns traded in a stock exchange. In our empirical investigation we study the MET of high frequency return time series of 20 highly capitalized stocks traded in New York Stock Exchange. Empirical results about this market confirm that the MET follows a power law with a pre-factor that depends on the specific stock chosen. This observation motivates us to first verify and then release some of the assumptions used in Ref. \cite{montero} therefore generalizing the model discussed in that paper. The theoretical generalization has been performed by introducing and solving a new two-state chain Markovian model able to both describe the quadratic scaling property of the MET and provide the data collapse of the MET stock pre-factor.

We show that a satisfactory data collapse of the MET is obtained when some degree of autocorrelation in the stock returns is introduced in the two-state chain Markovian model. We attempt to further improve the accuracy by extending the model to a three-state Markov chain for which we are still able to evaluate the MET. Nevertheless, empirical data show that the three-state model does not improve the quality of data collapse in the MET profiles of the different stocks although the theoretical curve shows a better agreement with the empirical data than that of the two-state Markov chain model.

The paper is organized as follows. In Sections \ref{Sect_Mean} and \ref{Sect_Discrete}, we discuss the MET behavior under the CTRW formalism  and a series of simplifying assumptions. In Sect.~\ref{Sect_Scaling} we empirically investigate the scaling and data collapse properties of highly capitalized stock data. Section~\ref{Sect_Correlation} relates the time correlation of stock return  with the absence of data collapse of MET observed in the previous section.  In Sect.~\ref{Sect_Markov} we introduce and solve a two-state and a three-state Markov chain model to describe the empirical MET observations. Conclusions are drawn in Sect.~\ref{Sect_Conclusions}.

\section{Mean exit time for i.i.d. processes}
\label{Sect_Mean}

\begin{figure}
  \begin{center}
  \includegraphics[width=0.45\textwidth,keepaspectratio=true]{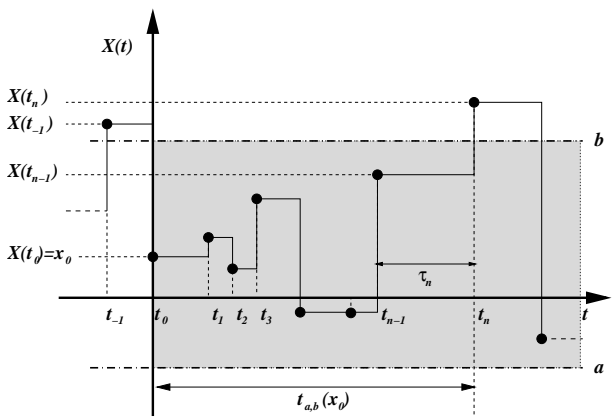} \caption{A sample trajectory of the $X(t)$ process along with the corresponding value of the random variable $t_{a,b}(x_0)$.} \label{model}
\end{center}
\end{figure}

In the most common version of the CTRW formalism a given random process $X(t)$ shows a series of random increments or jumps at random times $\cdots,t_{-1},t_0,t_1,t_2,\cdots,t_n,\cdots$ remaining  constant between these jumps. Therefore, after a given time interval $\tau_n=t_n-t_{n-1}$, the process experiences a random increment $\Delta X_n(\tau_n)=X(t_n)-X(t_{n-1})$ and the resulting trajectory consists of a series of steps as shown in Fig.~\ref{model}. Both waiting times $\tau_n$ and random jumps $\Delta X_n(\tau_n)$ are assumed to be independent and identically distributed (i.i.d.) random variables described by their probability density functions (pdfs) which we denote by $\psi(\tau)$ and $h(x)$ respectively. 

However, in the most general representation of the formalism, another function is needed to describe the time evolution of $X(t)$.  We denote this function by $\rho(x,\tau)$ which is the joint pdf of waiting times and jumps:
\begin{widetext}
\begin{equation}
                \rho(x,\tau)dxd\tau=
                \text{Prob}\{x<\Delta X_n\leq x+dx;\tau<\tau_n\leq \tau+d\tau\}. \label{rho0}
\end{equation}
\end{widetext}
Note that the functions $\psi(\tau)$ and $h(x)$ are the marginal probability density functions of $\rho(x,\tau)$. 
We refer the reader to Ref.~\cite{montrollweiss,weissllibre,general1,general2,general3,general4,general5,general6,masoliver1,masoliver2} for a more complete account of the CTRW formalism. 

In this paper we will apply the CTRW to study some aspects of the exit problem of financial time series. We will take as underlying random process $X(t)$ the logarithmic price $X(t_n)=\ln(S(t_n))$, where $S(t)$ is the stock price at time $t$. We specifically consider the problem of obtaining the mean exit time of $X(t)$ out of a given interval $[a,b]$. We assume that at certain reference time $t_0$, right after an event, the price has a known value $X(t_0)=x_0$, $x_0 \in [a,b]$. We focus our attention on a particular realization of the process and suppose that at certain time $t_n>t_0$ the process first leaves the interval (see Fig.~\ref{model}). We call the interval $t_n-t_0$ the exit time out of the region $[a,b]$ and denote it by $t_{a,b}(x_0)$. This quantity is a random variable since it depends on the particular trajectory of $X(t)$ chosen and the MET is simply the average $T_{a,b}(x_0)=\text{E}[t_{a,b}(x_0)]$.

The standard approach to exit time problems is based on the knowledge of the survival probability which is generally quite involved~\cite{weissrubin}. However, within the CTRW formalism one can assume that the events compose a series of independent and identically distributed two-dimensional random variables. Under such an assumption, some of us~\cite{montero} have recently shown that one can obtain the MET directly, without making use of the survival probability. In this framework the MET obeys the following integral equation~\cite{montero,footnote_ab}
\begin{equation}
T(x_0)=\text{E}[\tau] +\int_a^bh(x-x_0)T(x)dx,
\label{met_iid}
\end{equation}
where 
$$
\text{E}[\tau] =\int_0^{\infty}\psi(\tau') \tau' d\tau'
$$
is the mean waiting time between jumps. It is worth noticing that Eq.~(\ref{met_iid}) is still valid even when $\tau_n$ and $\Delta X_n$ are cross-correlated. In fact, in the case of an i.i.d. process the MET only depends on the pdfs of waiting times $\psi(\tau)$ and jumps $h(x)$, but it does not depend on the particular form of the joint pdf $\rho(x,\tau)$. However if we would remove the i.i.d. hypothesis we should specify a functional form for $\rho(x,\tau)$.

We now assume that returns increments are distributed according to an even pdf, $h(x)=h(-x)$, which also satisfies the following scaling condition,
\begin{equation}
h(x)=\frac{1}{\kappa}H\left(\frac{x}{\kappa}\right)
\label{h_scaled}
\end{equation}
where $\kappa$ is the scale of the fluctuations given by the standard deviation $\kappa$ of jumps, where $\kappa^2 =$ $\text{E}[\Delta X_n^2 - E[\Delta X_n]^2]$. The parameter $\kappa$ corresponds to the transaction-to-transaction volatility. Under these assumptions some of us showed that the MET out of a small region of size $L \equiv b-a\ll\kappa$ is~\cite{montero}
\begin{eqnarray}
T(a&+&L/2)=\text{E}[\tau] \left[1+2 H(0)\left(\frac{L}{2\kappa}\right)\right.\nonumber \\ 
&+&(H'(0^{+})+4 H(0)^2)\left(\frac{L}{2\kappa}\right)^2
+ \left. O\left(\frac{L^3}{\kappa^3}\right)\right],
\label{start}
\end{eqnarray}
where it is assumed that the return process is initially in the middle of the interval $[a,b]$. 

In Ref.~\cite{montero} some of us have applied the above result to the FX market of the U.S. dollar/Deutsche mark future price. One important conclusion there was that the quadratic growth of the MET is still a good approximation even for large intervals, i.e. $L\gg \kappa$. It is less clear nonetheless that the coefficients of the polynomial which we have obtained in Eq.~(\ref{start}) are those that better reproduce the global behavior of the MET. We refer the reader to Ref.~\cite{montero} for a more detailed discussion on this particular aspect of the problem.

\begin{figure}[htb]
\begin{center}
              \includegraphics[width=0.45\textwidth,keepaspectratio=true]{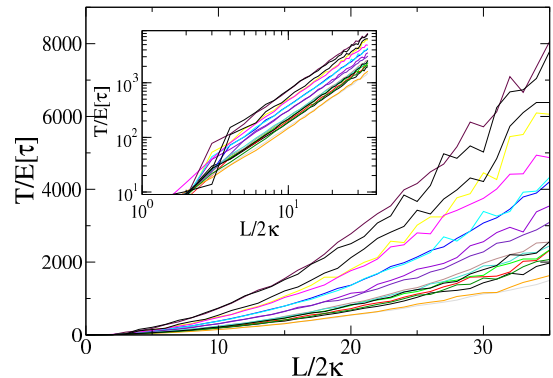}
              \caption{(Color online) Mean exit times as a function of the region size $L$ for the $20$ investigated stocks. The $x$ axis is scaled by $2 \kappa$, whereas the $y$ axis is scaled by the mean waiting time $\text{E}[\tau] $. The inset shows the same curves in a double logarithmic plot.} \label{startscaling}
\end{center}
\end{figure}

In this paper we analyze the scaling properties of the MET for $20$ highly capitalized stocks traded at the NYSE in the four year period 1995-1998 and spanning $1,011$ trading days.  Table~\ref{summary} shows the list of stocks and the relevant parameters. We have measured the MET for each stock and compared them in the scaled variables $T(a+L/2)/ \text{E}[\tau] $ and $L/2 \kappa$. If all previous hypotheses of the model were correct, and the function $H(u)$ is of universal nature, one should observe 
the same curve for all stocks, i.e. a data collapse, as well as a quadratic growth in $L$ at least for small $L$. However, Fig.~\ref{startscaling} shows that there is a considerable spread of the curves. although the parabolic shape is recovered in all cases not only for small intervals, as expected, but for the whole investigated range of $L/2\kappa$ .

\begin{table}[htb]
\caption{Summary statistics of the $20$ stocks we study for the period
  1995-1998.  The second column gives the number of transactions, the third one the standard deviation of jumps and the fourth one the mean waiting time, i.e. the mean time between two intraday consecutive transactions.}
\label{summary}
\begin{ruledtabular}
\begin{tabular}{l|ccc}
Ticker & Transactions & $\kappa$ ($\times 10^{-4}$)& $\text{E}[\tau] $ (seconds) \cr \hline 
AHP&   521,639  &  9.18    &44.2 \\  
AIG&   472,393  &  6.98    &49.2 \\
BMY&   573,397  &  7.30    &40.3 \\
CHV&   449,328  &  8.47    &51.5 \\
DD&    645,164  &  8.80    &36.0 \\
GE&  1,319,145  &  7.17    &18.7 \\
GTE&   512,581  & 13.10    &45.2 \\
HWP&   930,003  &  9.20    &25.0 \\
IBM& 1,072,395  &  5.42    &22.7 \\
JNJ&   728,686  &  8.75    &32.1 \\
KO&    784,357  &  9.11    &29.7 \\
MO&    971,700  & 10.11    &23.9 \\
MOB&   461,669  &  6.82    &50.0 \\
MRK&   971,842  &  7.34    &23.8 \\
PEP&   767,929  & 15.48    &30.3 \\
PFE& 1,003,518  &  7.46    &23.1 \\
PG&    679,601  &  7.36    &34.2 \\
T&   1,030,761  & 11.13    &22.6 \\
WMT&   565,946  & 15.80    &40.1 \\
XON&   674,412  &  7.08    &34.5 
\end{tabular}
\end{ruledtabular}
\end{table}

\section{A discrete state model}
\label{Sect_Discrete}

Our first objective is to understand why the quadratic term governs both the long and short range behavior of the MET, without a drastic change of the general model and its assumptions. In the present approach we develop a model for $h(x)$ based on the small-scale properties of the system. 

It is worth noticing that this approach is also used in the context of option pricing, when the fair price of a derivative product is obtained by making use of the binomial trees methodology, where it is assumed that the stock price makes a jump up or down with some  probability \cite{CRR}. Here we introduce the following symmetrical three-state discrete model:
\begin{equation}
                h(x)=Q\ \delta(x) + \frac{1-Q}{2}\left[\delta(x-c)+\delta(x+c)\right]. \label{2s_h}
\end{equation}
where $Q$ represents the probability that the price remains unchanged, and $c$ is the basic jump size \cite{footnote_grid}. By substituting this expression of $h(x)$ into Eq.~(\ref{met_iid}) we obtain 
\begin{equation*}
                 T(x_0)=\text{E}[\tau] +Q\ T(x_0) + \frac{1-Q}{2}\left[T(x_0+c)+T(x_0-c)\right],
\end{equation*}
with the convention that the term $T(x_0+c)$ only counts if $x_0\leq b-c$, and similarly that $T(x_0-c)$ only appears when $x_0\geq a+c$. Let us analyze these two boundary conditions in greater detail. In general, the limits of our interval can be expressed in the following form: 
\begin{eqnarray*}
                 a=x_0-(l+\varepsilon_a)c, \\ 
                 b=x_0+(m+\varepsilon_b)c,
\end{eqnarray*}
with $l,m \in \mathbb{N}$ and $\varepsilon_a,\varepsilon_b \in [0,1)$. However it  is easy to conclude that $T(x_0)$ can depend neither on $\varepsilon_a$ nor on $\varepsilon_b$. The only way of leaving the interval is by reaching the points $x=x_0-(l+1)c$ or $x=x_0+(m+1)c$, because $x=x_0-lc$, and $x=x_0+mc$ lay always inside the interval. Therefore we will not loose generality by setting $\varepsilon_a=\varepsilon_b=0$. After that, the length of the interval in the natural scale units of the problem is $N \equiv L/c=l+m$, and the use of the following notation $T_n \equiv T(a+nc)$, with $n\in \{0,1,\cdots,N-1,N\}$, arises in a natural way.

Summing up, for the discrete model given by Eq. (\ref{2s_h}) the MET out of the interval $[a,b]$ obeys the following set of difference equations
\begin{equation}
                T_n=\frac{\text{E}[\tau]}{1-Q}+\frac{1}{2} \left(T_{n+1}+T_{n-1} \right) \label{2s_met}
\end{equation}
$(n=0,1,2,\cdots,N)$ with boundary conditions: 
\begin{equation}
                T_{-1}=T_{N+1}=0. \label{2s_bc}
\end{equation}
The solution to Eqs. (\ref{2s_met})-(\ref{2s_bc}) is given by
\begin{equation}
                 T_n=\text{E}[\tilde\tau](n+1)(N+1-n),
\label{2s_solution}
\end{equation}
where the random variable $\tilde\tau$ is related to $\tau$ in such a way that 
\begin{equation}
                 \text{E}[\tilde\tau]=\frac{\text{E}[\tau]}{1-Q}. \label{tilde_tau}
\end{equation}
By repeating the above derivation leading to Eq. (\ref{2s_met}), one can show that the random variable $\tilde\tau$ represents the waiting time between jumps if one neglects zero-return trades, i.e. if one identifies the occurrence of a jump when $X(t)$ actually changes its value. Thus, for instance, if $\Delta X_{i-1}\ne 0$, $\Delta X_i=0$, and $\Delta X_{i+1}\ne 0$ with corresponding waiting times $\tau_{i-1}$, $\tau_i$, and $\tau_{i+1}$, we can replace the pair of events $(\Delta X_i, \Delta X_{i+1})$ with a single transaction of size $\Delta\tilde{X}_{j}=\Delta X_i+\Delta X_{i+1}=\Delta X_{i+1}$, taking a waiting time $\tilde{\tau}_j=\tau_{i}+\tau_{i+1}$. 

From Eq. (\ref{2s_solution}) we see that, for even values of $N$, the MET starting from the middle of the interval reads
\begin{equation}
                T_{N/2}=\text{E}[\tilde\tau]\left(1+\frac{N}{2}\right)^2.
\label{middle_even}
\end{equation}
Looking at the general solution given by Eq. (\ref{2s_solution}) and also at Eq. (\ref{middle_even}) we clearly observe a quadratic behavior of the MET as a function of $N$, that is to say, as a function of the length of the interval. Indeed, from Eq. (\ref{middle_even}) we have:
\begin{equation}
                \frac{T(a+L/2)}{\text{E}[\tilde\tau]}=
                              \left(1+\frac{L}{2\tilde\kappa}\right)^2. \label{met_iid_sol_exact_a}
\end{equation}
where $N=L/c$ and $\tilde\kappa^2=$ $\text{E}[\Delta\tilde{X}_n^2 - E[\Delta\tilde{X}_n]^2]$ $=c^2$. The same kind of scaling also holds in terms of the parameters of the three-state model. In fact, from Eq. (\ref{tilde_tau}) we get
\begin{equation}
               \frac{T(a+L/2)}{\text{E}[\tau]}=
                     \frac{1}{1-Q}\left(1+\frac{L\sqrt{1-Q}}{2\kappa}\right)^2. \label{met_iid_sol_exact}
\end{equation}
where $\kappa^2=$ $\text{E}[\Delta X_n^2-E[\Delta X_n]^2]$ $=(1-Q)\,c^2$. Hence, for large values of $L/\kappa$,
\begin{equation}
\frac{T(a+L/2)}{\text{E}[\tau] }\sim \left(\frac{L}{2\kappa}\right)^2.
\label{met_iid_sol_b}
\end{equation}

\section{Causes of absence of data collapse} \label{Sect_Scaling}

In the previous section we have shown that a simple discrete model for the jump distribution results in a quadratic growth of the MET valid for arbitrary values of the length of the interval and not only for small values of $L/\kappa$, as was the case of Eq. (\ref{start}). 

Unfortunately the discrete model does not properly account for the spread of the MET curves observed in Fig.~\ref{startscaling} when we consider different stocks. One could argue that this spread can be controlled through the parameter $Q$ appearing in Eq. (\ref{met_iid_sol_exact}) since $Q$ may distinguish one stock from another. However, as Eq. (\ref{met_iid_sol_b}) shows the MET is practically independent of $Q$ for large values of $L$ and the difference between stocks would disappear in this range of lengths. Nevertheless, we clearly see in Fig.~\ref{startscaling} that the spread between stocks does not tend to vanish but, even in some cases, it increases with $L$. 

In this Section we try to identify the possible reasons of the failure of the data collapse of our previous model. We revisit some of our assumptions and derive consequences with the aim of finding the most important feature that we are leaving aside. The final goal is to improve our description in the simplest possible way. Let us first summarize some potential causes for the lack of data collapse in the previous models:
\begin{enumerate}
\item The probability density $h(x)$ is different for different stocks. This would imply different mean exit time curves.

\item There is some dependency on the cross-correlation between waiting times and price returns. 

\item The time auto-correlation of waiting times should be included in the model.

\item The time auto-correlation of returns should be included in the model.
\end{enumerate}

\begin{figure}[htb]
\begin{center}
              \includegraphics[width=0.45\textwidth,keepaspectratio=true]{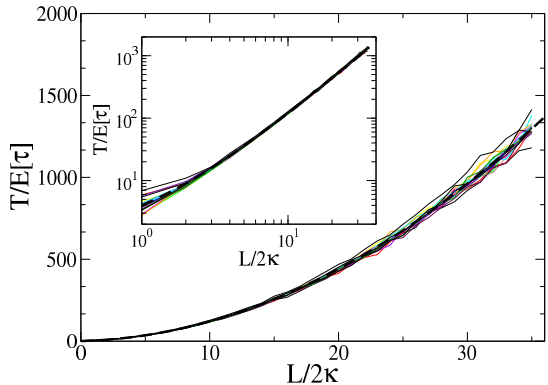}
              \caption{(Color online) Mean exit times as a function of the region size $L$ for the $20$ investigated time series after shuffling both waiting times and return increments. The $x$ axis is scaled by $2 \kappa$, whereas the $y$ axis is scaled by the mean waiting time $\rm{E}[\tau]$. The inset shows the same curves in a double logarithmic plot. The dashed curve represents the functional form $y=(1+x)^2$.} \label{shuffling2}
\end{center}
\end{figure}

We analyze the impact of these hypotheses on the empirical outcomes by performing shuffling experiments. Thus, in order to test the first hypothesis we shuffle independently the time series of $\Delta X_n$ and $\tau_n$ and we perform the mean exit time analysis on the shuffled time series. This shuffling destroys all the time- and cross- correlations but it preserves the shape of the pdfs $h(x)$ and $\psi(\tau)$ and therefore the values of the scaling parameter $\kappa$ and the average waiting time $\text{E}[\tau]$. Figure~\ref{shuffling2} shows a very good collapse indicating that the assumption of a master pdf for the returns of all the stocks is good working hypothesis and it is not the reason for the lack of collapse of Fig.~\ref{startscaling}. Moreover the MET curves are well fitted by the functional form $y=(1+x)^2$. 

\begin{figure}[htb]
\begin{center}
             \includegraphics[width=0.45\textwidth,keepaspectratio=true]{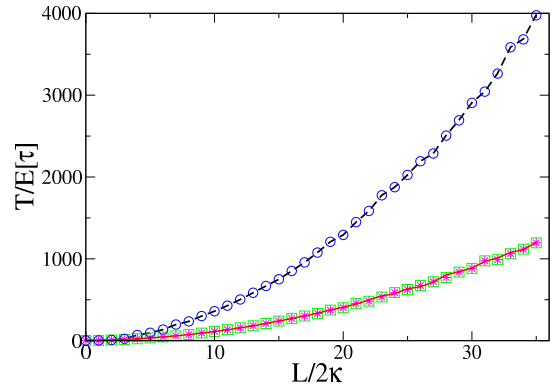}
             \caption{(Color online) Mean exit times as a function of the region size $L$ for General Electric obtained by performing four different shuffling experiments. The dashed (black) curve is the MET of the original data. The solid (red) line is the MET for data shuffled both in the waiting times and in the return increments. The squares (green) give the MET when one shuffles both $\Delta X_n$ and $\tau_n$ but preserving the association (cross correlation) between the two and circles (blue) give the MET when one shuffles only the waiting times. The star (magenta) give the MET when one shuffles only the returns.} \label{4tests}
\end{center}
\end{figure}

The other three hypotheses can be similarly tested by performing three different shuffling experiments. Specifically, hypothesis 2) can be tested by shuffling simultaneously the two series and preserving the cross-correlation between $\Delta X_n$ and $\tau_n$. Notice that even if Ref. \cite{montero} has shown that in the absence of autocorrelation of waiting times and returns (i.i.d. model) the MET is independent of the cross-correlations $\rho(x,\tau)$, in the general case $\rho(x,\tau)$ may play a role. Hypothesis 3) can be tested by shuffling only the series of waiting times and preserving the order in the series of returns. Finally, hypothesis 4) can be tested by shuffling only the time series of returns and preserving the order of the series of waiting times. Figure~\ref{4tests} shows the results of these shuffling experiments for the General Electric (GE) stock. The Figure shows that neglecting the autocorrelation function of waiting times does not change the MET curve. On the other hand, when one destroys the auto correlation of returns the MET (star) changes dramatically and becomes close to the one predicted by the i.i.d. model. 

By summarizing, in order to have a good model of the mean exit time one cannot neglect the correlation properties of price return, whereas the other correlations can be neglected as a first approximation.

\section{The correlation of return increments}
\label{Sect_Correlation}

What is the origin of correlation of returns? There are, in principle, two possible answers. One contribution comes from the linear autocorrelation of the increments of returns given by $\text{E}[\Delta X_n \Delta X_m]$. A second contribution is related to nonlinear properties which can be exemplified by the nonlinear correlation $\text{E}[|\Delta X_n||\Delta X_m|]$. 

\begin{figure}[htb]
\begin{center}
              \includegraphics[width=0.45\textwidth,keepaspectratio=true]{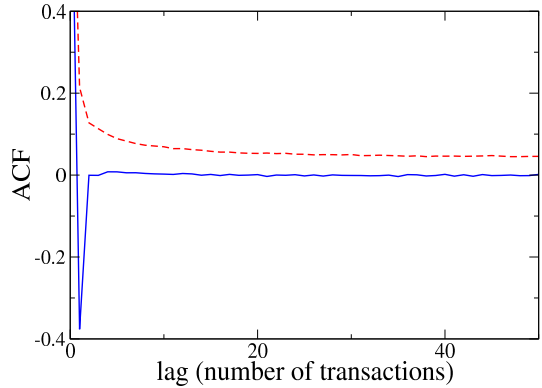}
              \caption{(Color online) The solid (blue) line shows the autocorrelation function of the time series of returns for the stock GE. The dashed (red) line shows the autocorrelation of the absolute value of returns for the same stock.}
\label{correlation}
\end{center}
\end{figure}

The first contribution can be easily evaluated by taking the linear autocorrelation function of the increments of price returns. Figure~\ref{correlation} shows this quantity for GE (solid line). In the Figure it is clear that for a lag of one trade the linear autocorrelation function is negative and significantly different from zero. This is a known effect of transaction prices which is due to the presence of a spread between the best bid and the best ask (the ``bid-ask bounce", see for example~\cite{campbell}). It is reasonable to assume that this short range correlation can be included in the model by using a Markov process.

The nonlinear correlation is related to the volatility and can be quantified by plotting the autocorrelation function of $|\Delta X_n|$. The result (dashed line) in Fig.~\ref{correlation} is a slowly decaying function indicating a long range correlation, which is probably not compatible with a Markovian model.

\begin{figure}[htb]
\begin{center}
              \includegraphics[width=0.45\textwidth,keepaspectratio=true]{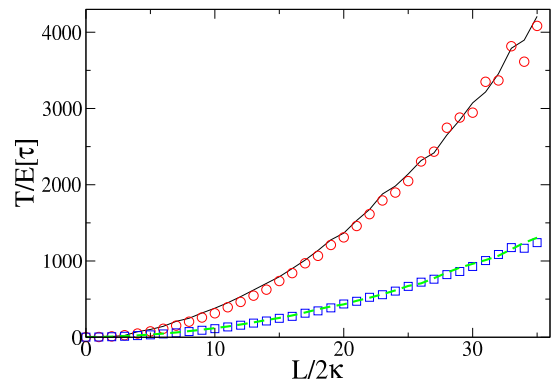}
              \caption{(Color online) Mean exit times obtained by performing shuffling experiments for the stock GE. The dashed (green) line refers to the real data shuffled both in time and in returns (the curve is also shown in Fig.~\ref{shuffling2}). The squares (blue) refer to the shuffled phase-randomized time series. The solid (black) curve refers to the original data (the curve is also shown in Fig.~\ref{startscaling}) and the circles (red) refer to the phase-randomized time series.} \label{surrogate}
\end{center}
\end{figure}

We can test the relative importance of the two contributions to the  correlation of returns by performing another shuffling experiment. We can obtain a surrogate time series with the same linear autocorrelation function but with an uncorrelated volatility (absolute value of return increments). The method (see for example Chapter 7 of Ref.~\cite{kantz}) consists in taking the Fourier transform of the original time series and then randomize its phases. Because of the Wiener-Kinchine theorem the linear autocorrelation of the surrogate time series is the same as the original, but the nonlinear correlation will be zero. Therefore, we have a time series with the same bid-ask bounce properties but with an uncorrelated volatility. One difficulty of the method is the fact that the pdf of the surrogate series will be in general different from the original one (unless the time series is Gaussian). Since we know that, in general, the MET depends on the return pdf, we should control that the distortion of the pdf introduced by the phase randomization is not critical in changing the properties of the MET. To this end we first compare the MET curve of the shuffled original time series to the shuffled phase-randomized time series. These two series are both i.i.d. but with different pdfs. Figure~\ref{surrogate} shows that the two METs (dashed line and squares) are very close, thus indicating that the distortion of the pdf introduced by the phase randomization is not critical for the MET properties.

We can now compare the METs of the original data to the ones for the phase-randomized time series. The two series have the same linear autocorrelation function, but the first one displays a clustered volatility whereas the randomized series does not. The two METs (see solid line and circles in Fig.~\ref{surrogate}) are again very close, thus showing that the most important time correlation contribution to the MET is the linear autocorrelation (bid-ask bounce), while the clustered volatility plays a minor role.

\section{Mean exit time for a Markov-chain model}
\label{Sect_Markov}

In order to incorporate the linear correlations of returns, we derive an integral equation for the mean exit time out of a given interval when the driving process is a continuous time random walk with memory. In particular, we will consider  Markov processes within the CTRW framework. Such models lead to the following joint conditional probability density function:
\begin{widetext}
\begin{equation}
                \rho(x,\tau|x^{\prime},\tau^{\prime})dxd\tau=\text{Prob}
                    \{x<\Delta X_n\leq x+dx;\tau<\tau_n\leq \tau+d\tau|
                      \Delta X_{n-1}=x^{\prime};\tau_{n-1}=\tau^{\prime}\}. 
\label{rho_corr1}
\end{equation}
\end{widetext}
In this Markovian case the conditional MET, $T(x_0|\Delta X_0,\tau_0)$, will also depend on both the magnitude of the previous jump $\Delta X_0=x_0-x_{-1}$ and its sojourn time $\tau_0=t_0-t_{-1}$ (see Fig. \ref{model}). Now, and contrary to Eq. (\ref{met_iid}), the integral equation for the conditional MET depends on the complete joint probability density function. It reads:
\begin{widetext}
\begin{equation}
                T(x_0|\Delta X_0,\tau_0)=\text{E}\left[\tau|\Delta X_0,\tau_0\right]+
                                         \int_0^{\infty} d\tau \int_{a}^{b} 
                                             \rho(x-x_0,\tau|\Delta X_0,\tau_0)T(x|\Delta X,\tau)dx, 
\label{MET_Markovian0}
\end{equation}
\end{widetext}
where $\Delta X=x-x_{0}$ and $\tau=t-t_{0}$. The level of complexity of Eq. (\ref{MET_Markovian0}) can be considerably reduced by noting that, as we have shown in Sect. \ref{Sect_Correlation}, it is possible to remove the correlation between consecutive waiting times without affecting the MET. Therefore, we will assume that the correlation involving waiting times is negligible, and that all relevant information we have to consider when dealing with the $n$-th event is the magnitude of the previous change. In such a case instead of Eq. (\ref{rho_corr1}) we write 
\begin{widetext}
\begin{equation}
                \rho(x,\tau|x^{\prime})dxd\tau=\text{Prob}
                     \{x<\Delta X_n\leq x+dx;\tau<\tau_n\leq \tau+d\tau|\Delta X_{n-1}=x^{\prime}\}. \label{rho_corr2}
\end{equation}
\end{widetext}
Hence, the integral equation for the MET is simpler, because on the right hand side of Eq. (\ref{MET_Markovian0}) we can perform the integral over time. We thus obtain
\begin{eqnarray}
T(x_0|\Delta X_0)&=&\text{E}\left[\tau|\Delta X_0\right]
\label{MET_Markovian} \nonumber \\
&+&\int_{a}^{b} h(x-x_0|\Delta X_0)T(x|\Delta X) dx.
\end{eqnarray}
In this case the MET only depends on the marginal probability density function of the return increments, $h(x|\Delta X_0)$,
\begin{equation*}
h(x|\Delta X_0) = \int_0^{\infty} \rho(x,\tau|\Delta X_0) d\tau,
\end{equation*}
and on the conditional expectation of waiting times $\text{E}\left[\tau|\Delta X_0\right]$ which has to be evaluated through the marginal pdf, $\psi(\tau|\Delta X_0)$,    
\begin{equation*}
\psi(\tau|\Delta X_0) = \int_{-\infty}^{\infty} \rho(x,\tau|\Delta X_0) dx.
\end{equation*}
We finally observe that although $x_0 \in [a,b]$, we let  $\Delta X_0$ to be any real number.

\subsection{A two-state Markov chain model}
\label{Sect_Memory2}

In order to solve Eq. (\ref{MET_Markovian}) and obtain explicit expressions for the MET that can be compared with empirical data, we follow the same approach of Sect. \ref{Sect_Discrete} and choose a discrete model for $h(x|\Delta X_0)$. At this point we can opt for a two-state model in which, at any time step, returns can only go up and down a fixed quantity $c$, or for a three-state model where in addition the return increment can be zero. We have shown in Sect. \ref{Sect_Discrete} that for an i.i.d. process both alternatives are equivalent. In the case of a Markovian process the equivalence is not complete. As we will see below, the final expressions obtained for the unconditional MET are slightly different although, for large values of $L$, the leading term is the same in both cases.

Let us start with a two-state Markov chain model. In the symmetrical case in which up and down movements are equally likely, the conditional pdf for return increments is
\begin{equation}
h(x|y) = \frac{c+ry}{2c} \delta(x-c)+\frac{c-ry}{2c} \delta(x+c),
\label{h(x|y)}
\end{equation}
where $r$ is the correlation between the magnitude of two consecutive jumps:
\begin{equation}
r\equiv\frac{\text{Cov}[\Delta \tilde{X}_n,\Delta \tilde{X}_{n-1}]}
{\sqrt{\text{Var}[\Delta \tilde{X}_n]\text{Var}[\Delta \tilde{X}_{n-1}]}}. 
\label{r}
\end{equation}
From Eq. (\ref{h(x|y)}) we see that the squared volatility,
$$
\tilde{\kappa}^2(y)\equiv\int_{-\infty}^{\infty}x^2h(x|y)dx=c^2,
$$
is independent of $y$. By substituting Eq. (\ref{h(x|y)}) into Eq.~(\ref{MET_Markovian}) we get the following difference equation for the MET:
\begin{eqnarray}
T(x_0|\Delta \tilde{X}_0)&=&\text{E}[\tilde{\tau}|\Delta \tilde{X_0}]+\frac{c+r\Delta \tilde{X_0}}{2c} T(x_0+c|c) \nonumber\\
&+&\frac{c-r\Delta \tilde{X_0}}{2c} T(x_0-c|-c),
\label{MET_Discrete}
\end{eqnarray}
where $\Delta \tilde{X_0}=\pm c$, 
$T(x_0+c|c)=0$ if $x_0>b-c$ and $T(x_0-c|-c)=0$  if  $x_0<a+c$. We extend the notation introduced in 
Sect.~\ref{Sect_Discrete} and define
\begin{equation}
T_{n,n\mp 1}\equiv T(x_0=a+nc | \Delta \tilde{X}_0= \pm c).
\label{t_n_def}\
\end{equation}
Now, Eq.~(\ref{MET_Discrete}) is equivalent to the following set of recurrence equations:
\begin{eqnarray}
T_{n,n - 1}&=&\text{E}[\tilde{\tau}]+\frac{1+r}{2} T_{n+1,n}+\frac{1-r}{2} T_{n-1,n} \label{set_1},\\
T_{n,n + 1}&=&\text{E}[\tilde{\tau}]+\frac{1-r}{2} T_{n+1,n}+\frac{1+r}{2} T_{n-1,n},\label{set_2}
\end{eqnarray}
($n=0,1,\cdots,N$) with boundary conditions: 
\begin{equation}
T_{-1,0}=T_{N+1,N}=0.
\label{bc_set}
\end{equation}
Note that in writing Eqs. (\ref{set_1})-(\ref{set_2}) we have set $\text{E}[\tilde{\tau}|\pm c]=\text{E}[\tilde{\tau}]$  which is consistent with the assumed symmetry between up and down movements (see also Eq. (\ref{uncond_met_def}) below). The solution to problem 
(\ref{set_1})-(\ref{bc_set}) reads
\begin{eqnarray*}
T_{n,n - 1}&=&\text{E}[\tilde{\tau}](N+1-n)\left[1+n \frac{1-r}{1+r}\right],\\
T_{n,n + 1}&=&\text{E}[\tilde{\tau}](n+1)\left[1+(N-n) \frac{1-r}{1+r}\right].
\end{eqnarray*}

The quantity of interest for our analysis is the unconditional MET $T_n$, which is related to $T_{n,n \pm 1}$ by 
\begin{equation}
T_n=\frac{1}{2}\left(T_{n,n - 1}+T_{n,n + 1}\right),
\label{uncond_met_def}
\end{equation}
that is,
\begin{equation}
T_n=\text{E}[\tilde{\tau}]\left[1+\frac{N}{1+r}+\frac{1-r}{1+r}n(N-n)\right].
\label{uncond_met}
\end{equation}
Thus the MET starting from the middle of the interval reads
\begin{equation}
T_{N/2}=\text{E}[\tilde{\tau}]\left[\frac{2r}{1+r}\left(1+\frac{N}{2}\right)+\frac{1-r}{1+r}\left(1+\frac{N}{2}\right)^2\right].
\label{met_middle}
\end{equation}
In this case, and contrary to the i.i.d. case given in 
Eq. (\ref{middle_even}), the MET $T_{N/2}$ is not a perfect quadratic expression. However, taking into account that $N=L/c$ and $\tilde\kappa=c$ we have
\begin{equation}
\frac{T(a+L/2)}{\text{E}[\tilde{\tau}]}=\frac{2r}{1+r}
\left(1+\frac{L}{2\tilde{\kappa}}\right)+\frac{1-r}{1+r}
\left(1+\frac{L}{2\tilde{\kappa}}\right)^2,
\label{met_middle_2}
\end{equation}
and for large values of $L/\tilde\kappa$ we recover the expected quadratic behavior in the leading term:
\begin{equation}
\frac{T(a+L/2)}{\text{E}[\tilde{\tau}]}\sim\frac{1-r}{1+r}\left(1+\frac{L}{2\tilde{\kappa}}\right)^2.
\label{twostates}
\end{equation}
Note that the value of $r$ depends on each particular stock. Therefore, the scaled MET defined as:
\begin{equation}
T_{sc}(L)\equiv\left(\frac{1+r}{1-r}\right)
\frac{T(a+L/2)}{\text{E}[\tilde{\tau}]},
\label{t_rs}
\end{equation}
tends, for increasing values of $L$, to a quadratic function of the interval length which is independent of the particular stock chosen.

\begin{figure}[htb]
\begin{center}
          \includegraphics[width=0.45\textwidth,keepaspectratio=true]{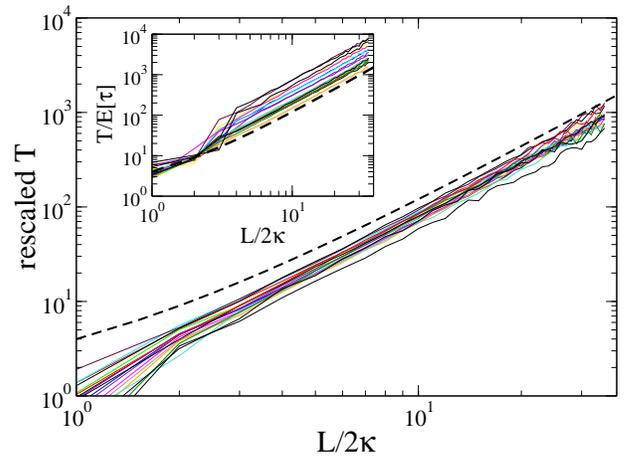}
          \caption{(Color online) Comparison of the MET curves for the original i.i.d. model and the two-state Markov chain model. In this figure we have scaled the MET according to Eq.~(\ref{twostates}) of the two-state Markovian model. The dashed (black) line is the parabolic curve $y=(1+x)^2$. The inset is the same shown in Fig.~\ref{startscaling} for the i.i.d. process.} \label{twostatesfig}
\end{center}
\end{figure}

In Fig.~\ref{twostatesfig} we compare the MET curves for the data scaled in the original way (inset), as given in 
Eq.~(\ref{met_iid_sol_exact_a}), and for the data scaled according to Eq.~(\ref{t_rs}). The inset of 
Figure~\ref{twostatesfig} is essentially the same as Fig.~\ref{startscaling}. Figure~\ref{twostatesfig} shows that the scaling of Eq.~(\ref{twostates}) gives a significant improvement with respect to the original one. We note that the scaled curves are systematically below the curve $y=(1+x)^2$, probably because some correlation is not taken correctly into account by the two-state model. We will now attempt to correct this bias by introducing a three-state model.

\subsection{A three-state Markov chain model}
\label{Sect_Memory3}

One can argue that the two-state model just developed would need an improvement in order to include zero-return transactions, i.e.  those with $\Delta X_n=0$. In Sect. \ref{Sect_Discrete} we have shown that for the i.i.d. process the inclusion of a third possible state is completely equivalent to a two-state (up and down) model after redefining the mean waiting time and the volatility $\kappa$ by including the probability $Q$ of zero-return transactions. However, when memory is present, as is now the case, this equivalence is not complete. We thus outline a discrete three-state Markov chain model. 

The Markov-chain model is now characterized by the following transition matrix:
\begin{equation}
\mathbf{T}=\left(\begin{array}{ccc}
 P(-|-)&P(-|0)&P(-|+)\\
 P(0|-)&P(0|0)&P(0|+)\\
 P(+|-)&P(+|0)&P(+|+)\\
 \end{array}\right),
 \label{transmat1}
\end{equation}
where $P(-|-)\equiv \text{Prob}\{\Delta X_n=-c|\Delta X_{n-1}=-c\}$ and similar definitions for the rest of the matrix elements.
Since we are also assuming that the process is symmetrical for positive and negative returns, the transition matrix $\mathbf{T}$ can be written in the following form:
\begin{equation}
\mathbf{T}=\left(\begin{array}{ccc}
\frac{1+r}{2}p&\frac{1-q}{2}&\frac{1-r}{2}p \\
1-p&q&1-p \\
 \frac{1-r}{2}p&\frac{1-q}{2}&\frac{1+r}{2}p \\
 \end{array}\right),
\label{transmat2}
\end{equation} 
where $q=P(0|0)$ is the probability for trapping,  
\begin{equation*}
p=P(-|-)+P(+|-)=P(+|+)+P(-|+)
\end{equation*}
and $r$ measures the strength of the persistence:
\begin{equation*}
r=\frac{P(-|-)-P(+|-)}{P(-|-)+P(+|-)}=
\frac{P(+|+)-P(-|+)}{P(+|+)+P(-|+)}.
\end{equation*}
Note that the first order autocorrelation coefficient, defined as on the right-hand side of Eq. (\ref{r}) is now given by ({\it cf.} Eq. (\ref{r}))
\begin{eqnarray}
             \frac{\text{Cov}[\Delta {X}_n,\Delta {X}_{n-1}]}
               {\sqrt{\text{Var}[\Delta {X}_n]\text{Var}[\Delta {X}_{n-1}]}}=pr. \label{pr_estimate}
\end{eqnarray}
We also need to specify the unconditional probabilities of each state: 
\begin{equation}
 {\bf P}=\left(\begin{array}{c}
 P(-)\\
 P(0)\\
 P(+)\\
 \end{array}\right)\equiv
  \left(\begin{array}{c}
\frac{1-Q}{2}\\
Q\\
\frac{1-Q}{2}\\
 \end{array}\right).
\end{equation}
However, in this case $Q$ is not an independent parameter as was the case of the i.i.d. model ({\it cf} Sect. \ref{Sect_Discrete}). Indeed, using the total probability formula and taking into account 
the values of the transition matrix $\mathbf{T}$ and the vector ${\bf P}$, we get 
\begin{equation*}
p=1-\frac{1-q}{1-Q}Q.
\end{equation*}
Observe that when $q=Q$ we have $p=1-Q$ and there is no trapping in the value of the random process.

Now the pdf of the returns reads
\begin{eqnarray}
h(x|y)&=&a(y)\delta(x)\label{h_three_states}\\
&+&\frac{(1-a(y))}{2c}\left[\left(c+ry\right)\delta(x-c)+\left(c-ry\right)\delta(x+c)\right],
\nonumber
\end{eqnarray}
where
$$
a(y)=\begin{cases} q, & \text{if $y=0$,}\\
            1-p, &\text{if $y\neq 0$.} 
      \end{cases} 
$$            
In this case the integral equation (\ref{MET_Markovian}) is equivalent to the following set of difference equations for $T_{n,n}$ and $T_{n,n \pm 1}$:
\begin{eqnarray}
T_{n,n - 1}&=&\text{E}[\tau|c]+ (1-p)T_{n,n}\nonumber\\
&+&p\left[\frac{1+r}{2} T_{n+1,n}+\frac{1-r}{2} T_{n-1,n}\right] 
\label{3set_1}
\end{eqnarray}
\begin{eqnarray}
\hspace*{-1.0em}T_{n,n}=\text{E}[\tau|0]&+& q T_{n,n}\nonumber\\
&+&\frac{1-q}{2}\left[T_{n+1,n}+T_{n-1,n}\right] 
\label{3set_2}
\end{eqnarray}
\begin{eqnarray}
T_{n,n + 1}&=&\text{E}[\tau|c]+ (1-p)T_{n,n}\nonumber\\
&+&p\left[\frac{1-r}{2} T_{n+1,n}+\frac{1+r}{2} T_{n-1,n}\right] 
\label{3set_3}
\end{eqnarray}
where $T_{n,n \pm 1}$ are defined as in Eq. (\ref{t_n_def}) and 
$$
T_{n,n}=T(x_0=a+nc|\Delta X_0=0).
$$
In writing Eqs. (\ref{3set_1}) and (\ref{3set_3}) we have taken into account the symmetry 
$$
\text{E}[\tau|+c]=\text{E}[\tau|-c]\equiv\text{E}[\tau|c].
$$ 

\begin{figure}[htb]
\begin{center}
               \includegraphics[width=0.5\textwidth,keepaspectratio=true]{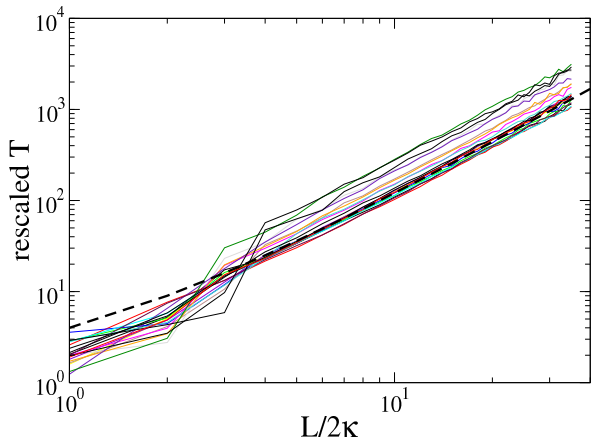}
               \caption{(Color online) MET curves scaled according to Eq. (\ref{first_manuscript}) of the three-state Markovian model. The quantity $p\,r$ has been estimated by using Eq. (\ref{pr_estimate}). The dashed (black) line represents the curve $y=(1+x)^2$.} \label{threestatesfig}
\end{center}
\end{figure}

Finally the solution to Eqs. (\ref{3set_1})-(\ref{3set_3}) with boundary conditions 
$$
T_{-1,0}=T_{N+1,N}=0
$$
({\it cf} Eq. (\ref{bc_set})) reads 
\begin{equation}
\hspace*{-3.5em} T_{n,n - 1}=\frac{\text{E}[\tau]}{1-Q}(N+1-n)\left[1+n\frac{1-pr}{1+pr}\right]
\label{3solution_1}
\end{equation}
\begin{eqnarray}
T_{n,n}=\frac{\text{E}[\tau|0]}{1-q}
&+&\frac{\text{E}[\tau]}{2(1-Q)}
\Biggl[\frac{1-pr}{1+pr}\Bigl[(N-n)(n+1)\nonumber\\
&+& n(N-n+1)\Bigr]+N\Biggr]
\label{3solution_2}
\end{eqnarray}
\begin{equation}
\hspace*{-3.5em} T_{n,n+1}=
\frac{\text{E}[\tau]}{1-Q}(n+1)\left[1+(N-n)\frac{1-pr}{1+pr}\right],
\label{3solution_3}
\end{equation}
where $\text{E}[\tau]$ is the (unconditional) mean waiting time which is related to $\text{E}[\tau|c]$ and $\text{E}[\tau|0]$ by
$$
\text{E}[\tau]=Q\text{E}[\tau|0]+(1-Q)\text{E}[\tau|c].
$$
In terms of $T_{n,n\pm 1}$ and $T_{n,n}$ the unconditional MET $T_n$ is given by
$$
T_{n}=QT_{n,n}+\left(\frac{1-Q}{2}\right)(T_{n,n-1}+T_{n,n+1}),
$$
and starting from the center of the interval we explicitly have
\begin{eqnarray}
T_{N/2}&=&\frac{\text{E}[\tau] }{1-Q}\left[\frac{2pr}{1+pr}\left(1+\frac{N}{2}\right)+\frac{1-pr}{1+pr}\left(1+\frac{N}{2}\right)^2\right]\nonumber\\
&+& Q\left[\frac{E[\tau|0]}{1-q}-\frac{E[\tau]}{1-Q}\right].
\label{met_three_states}
\end{eqnarray}
The main difference between this expression and Eq. (\ref{met_middle}) is the final constant term that accounts for the two kind of trapping that the system may experience: the probabilistic one, $q\neq Q$, and the temporal one, $E[\tau|0]\neq E[\tau]$. By using the fact that $\kappa^2=(1-Q)\,c^2$, the leading term is again of the form:
\begin{eqnarray}
      && \frac{T(a+L/2)}{\text{E}[\tau]}\sim
         \frac{1-pr}{1+pr}\left(\frac{L}{2\kappa}\right)^2. \label{first_manuscript}
\end{eqnarray}
In Fig.~\ref{threestatesfig} we show the MET curves for the data scaled according to 
Eq.~(\ref{first_manuscript}). It is worth noting that the quantity $p\,r$ appearing in Eq. (\ref{first_manuscript}) can be estimated in two different ways. One could separately compute $p$, the probability of a change in the price return provided a previous change, and $r$ which is the strength of the persistence. Alternatively one can estimate directly the quantity $p\,r$ by using Eq. (\ref{pr_estimate}). This second approach has the advantage of reducing the dependence of the estimates from the specific details of the model. This is the reason why in the curves shown in Fig. ~\ref{threestatesfig} the quantity $p\,r$ has been estimated by using Eq. (\ref{pr_estimate}). The scaling shown in Figure~\ref{threestatesfig} is not satisfactory, we would even say that it is less satisfactory than the scaling corresponding to the two-state model which has been shown in Fig.~\ref{twostatesfig}. However, it is worth noting that the rescaled curves are not systematically below the parabolic curve $(1+x)^2$ as it was in the case of the two-state model.

\section{Conclusions} \label{Sect_Conclusions}

This paper presents theoretical and empirical results about the MET of financial time series. Specifically the scaling property of the MET as a function of the size $L$ has been confirmed to follow a quadratic law for a number of stock price time series. We empirically verify that the quadratic scaling law has associated a pre-factor which is specific to the analyzed stock. We have performed a series of tests to determine which kind of correlation are responsible for this dependence. It turned out that the main contribution is associated with the linear autocorrelation property of stock returns.

We have therefore introduced and solved analytically both a two-state and a three-state Markov chain models. The analytical results obtained through the two-state model allow us to get a quite satisfactory data collapse of the 20 MET profiles into a single parabolic curve as predicted by the model. However, this parabolic curve appears to be systematically above real data, that is, the model overstimates the mean exit time.

We have been able to solve a three-state Markov chain model as well. Unfortunately this more detailed model does not provide an improvement on data collapse. The main advantage of this generalization is that the MET provided by the model lies close to the empirical curves. In other words, the three-state model does not overestimate the METs. 

We do not have a convincing explanation for this observation but only some indications. Specifically, we have seen that the symmetries assumed in the three state model are not present in some empirical transition matrices \cite{transition}. Perhaps this assumption prevents the data from a convincing data collapse and the system would perform a better collapse in a model taking into account a certain degree of asymmetry in the Markovian transition matrix. However, obtaining the analytical solution for this more general case seems to be very involved and it has been left for future research.

In conclusion, the MET and the search of a data collapse in the MET curves of stock prices provide a good occasion for testing the underlying hypothesis characterizing the return  dynamics and for the improvement of the CTRW models describing this phenomenon. We believe we have detected the essential ingredients to be accounted for a feasible model within the CTRW. We hope that the results may provide a certain insight on the way to bridge the gap between the description of the stochastic dynamics at very short time horizons with that of longer time scales 
\cite{campbell,ohara,gabaix1,gabaix2,gabaix3,gabaix4,gabaix5}.

\acknowledgments The authors acknowledge support from the ESF project ``Cost Action P10 Physics of Risk''. M.M., J.P. and J.M. acknowledge partial support from Direcci\'on General de Investigaci\'on under contract No. BFM2003-04574 and by Generalitat de Catalunya under contract No. 2001 SGR-00061.  F.L., S.M. and R.N.M  acknowledge support from the research project MIUR-FIRB RBNE01CW3M ``Cellular Self-Organizing nets and chaotic nonlinear dynamics to model and control complex system'', from the research project MIUR 449/97 ``High frequency dynamics in financial markets" and from the European Union STREP project n. 012911 ``Human behavior through dynamics of complex social networks: an interdisciplinary approach.''.


\begin{thebibliography}{99}
\bibitem{footnote_ab} If there is no confusion and in order to keep notation simple, we will drop the subscript and write $T(x_0)$ for the MET out of $[a,b]$.
\bibitem{transition} For instance, the General Electric transition matrix has the following elements: $P(-|-)=0.05$, $P(-|0)=0.14$, $P(-|+)=0.34$, $P(0|-)=0.53$, $P(0|0)=0.73$, $P(0|+)=0.62$, $P(+|-)=0.42$, $P(+|0)=0.13$, and $P(+|+)=0.04$. Note that some symmetry assumptions on the transition matrix of Eqs. (\ref{transmat1})-(\ref{transmat2}) are approximately satisfied: $P(-|-)\simeq P(+|+)$ and $P(-|0)\simeq P(+|0)$ while others are not: $P(+|-)\neq P(-|+)$ and $P(0|-)\neq P(0|+)$.
\end{thebibliography}
\end{document}